\RequirePackage{fix-cm}
\documentclass[twocolumn]{svjour3}          
\smartqed  
\usepackage{graphicx}
\usepackage{siunitx}
\usepackage{todonotes}
\usepackage{hyperref}
\usepackage{amsmath}
\journalname{Eur. Phys. J. A}
\begin{document}

\title{Direct TPE Measurement via $e^+p/e^-p$ Scattering at low $\varepsilon$ in Hall A
}


\author{Ethan Cline         \and
        Jan C. Bernauer  \and 
        Axel Schmidt
}


\institute{Ethan~Cline \and Jan~C.~Bernauer  \at Center for Frontiers in Nuclear Science, Stony Brook University, Stony Book, NY, USA
          \and
          Jan~C.~Bernauer \at Riken BNL Research Center, Upton, NY, USA 
        \and
          Axel~Schmidt \at   The  George  Washington  University,  Washington,  DC, USA \label{gwu}
}

\date{Received: date / Accepted: date}

\maketitle

\begin{abstract}
The proton elastic form factor ratio can be measured either via Rosenbluth separation in an experiment with unpolarized beam and target, or via the use of polarization degrees of freedom. However, data produced by these two approaches show a discrepancy, increasing with $Q^2$. The proposed explanation of this discrepancy – two-photon exchange – has been tested recently by three experiments. The results support the existence of a small two-photon exchange effect but cannot establish that theoretical treatments at the measured momentum transfers are valid. At larger momentum transfers, theory remains untested, and without further data, it is impossible to resolve the discrepancy. A positron beam at Jefferson Lab allows us to directly measure two-photon exchange over an extended $Q^2$ and $\epsilon$ range with high precision. With this, we can validate whether the effect reconciles the form factor ratio measurements, and test several theoretical approaches, valid in different parts of the tested $Q^2$ range. In this proposal, we describe a measurement program in Hall A that combines the Super BigBite, BigBite, and High Resolution Spectrometers to directly measure the two-photon effect. Though the limited beam current of the positron beam will restrict the kinematic reach, this measurement will have very small systematic uncertainties, making it a clean probe of two photon exchange.
\end{abstract}

\section*{Introduction}

\begin{figure*}[htb]
  \centerline{\includegraphics[width=\linewidth]{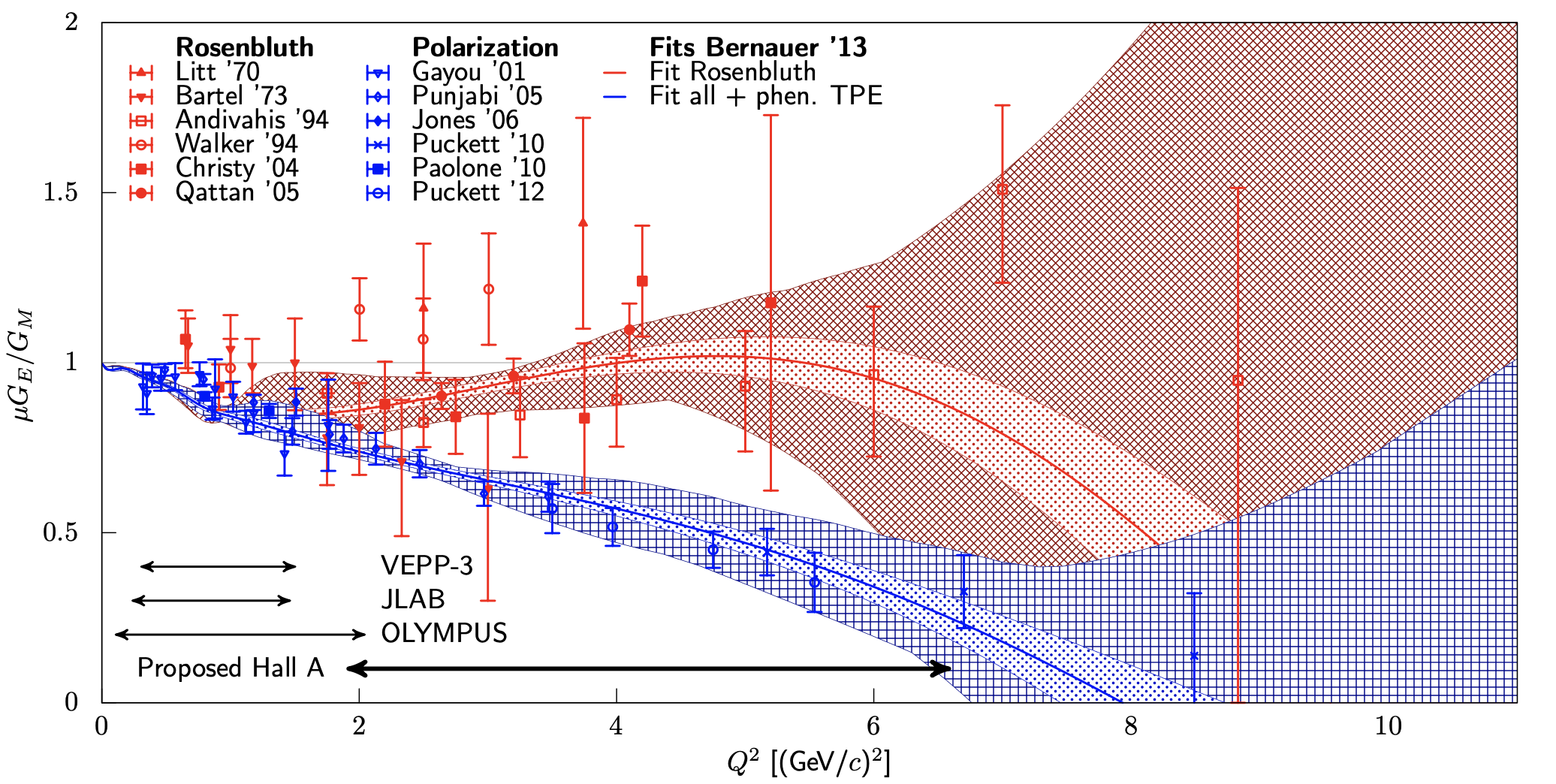}}
  \caption{\label{figratio}The proton form factor ratio $\mu G_E/G_M$, as determined via Rosenbluth-type (red points, from \cite{litt,bartel,andivahis,walker,christy,qattan}) and polarization-type (blue points, from \cite{gayou,punjabi,jones,puckett10,paolone,puckett12}) experiments. While the former indicate a ratio close to 1, the latter show a distinct linear fall-off. Curves are from a phenomenological fit \cite{Bernauer:2013tpr}, to either the Rosenbluth-type world data set alone (red curves) or to all data, then including a phenomenological two-photon-exchange model. Note that the red fit curve includes the Mainz data not shown here. We also indicate the coverage of earlier experiments as well as of the experiment described below.}
\end{figure*}

Historically the elastic form factors of the proton have been probed via unpolarized electron-proton scattering. The cross sections were measured  over a large range of four-momentum transfer squared, $Q^2$, and the form factors were extracted via the so-called Rosenbluth separation. It was found that the form factor ratio $\mu G_E/G_M$ is constant and consistent with unity over the measured range of data. Here $\mu$ is the magnetic moment of the proton and $G_E$ ($G_M$) is the Sach's electric (magnetic) form factor of the proton. Somewhat more recently, the ratio of the form factors was measured using polarized electron beams, with different systematics and increased precision. However, the polarized results indicate a roughly linear fall-off of the ratio, decreasing with increasing $Q^2$. A summary of the different experimental measurements is compiled in Fig.\ \ref{figratio}. The two data sets are inconsistent with each other, indicating that the true proton form factors are unknown, as either method could be introducing errors into the extraction. The resolution of this ``form factor ratio puzzle" is crucial to advance our knowledge of the proton electromagnetic form factors. As this puzzle seems to increase with $Q^2$, it is critical to determine the resolution, especially in light of JLab's new high $Q^2$ measurement program.

The differences observed between the polarized and unpolarized methods have been attributed to two-photon exchange (TPE) effects \cite{guichon03,carlson07,arrington11,Afanasev:2017gsk}, which are much more important in the Rosenbluth method than in the polarization transfer method, where in the ratio of longitudinal to transverse polarization cross-sections the TPE partially cancels out.  TPE corresponds to the group of diagrams where two photon lines connect the lepton and proton. The diagrams are typically separated into two cases, ``soft" and ``hard", although it is important to note that this distinction is arbitrary, and depends on the specific prescription being employed. The so-called ``soft'' case, when one of the photons has negligible momentum, is included in the standard radiative corrections, like ref.\ \cite{MoTsai,Maximon2000}, to cancel infrared divergences from other diagrams. The ``hard'' part, where both photons can carry considerable momentum, is not.

It is obviously important to study this proposed solution to the discrepancy with experiments that have sensitivity to two-photon contributions. 
The most straightforward process to evaluate two-photon contribution is the measurement of the ratio of elastic $e^+p/e^-p$ scattering, which in leading order is given by the expression:
\begin{equation}
    R_{2\gamma} = 1 - 2\delta_{\gamma\gamma} \, .
\end{equation}

\begin{figure}[t]
  \centerline{\includegraphics[width=\linewidth]{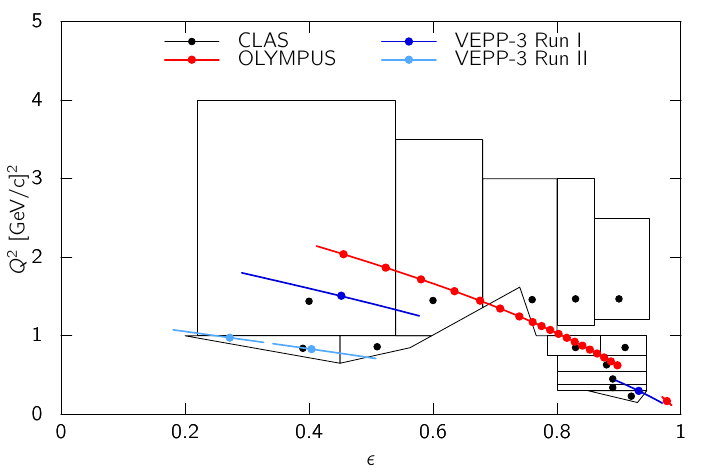}}
  \caption{ Kinematics covered by the three recent experiments to measure the TPE contribution to the elastic ep cross section. The beam energy in the CLAS experiment was not fixed. The black polygons signify the phase space of data projected to the black points.}
  \label{reach}
\end{figure}

The ratio of $e^+p/e^-p$ deviates from 1 due to the TPE, as some radiative diagrams depend on lepton charge to the third power, i.e. change sign with the lepton charge sign, which leads to differences in the respective cross section. Here we define $\delta_{\gamma\gamma}$ as the TPE.
Several experiments have recently been carried out to measure the two-photon exchange contribution in elastic scattering: the VEPP-3 experiment at Novosibirsk \cite{Rachek:2014fam}, the CLAS experiment at Jefferson Lab \cite{Rimal:2016toz,Adikaram:2014ykv,moteabbed13}, and the OLYMPUS experiment at DESY \cite{Henderson:2016dea}. 
The kinematic reach of these experiments was limited, however, as seen in Fig.~\ref{reach}.

The current status can be summarized as such:
\begin{itemize}
\item TPE exists, but is small in the covered region.
\item Hadronic theoretical calculations, supposed to be valid in this kinematic regime, might not be good enough yet.
\item Calculations based on GPDs, valid at higher $Q^2$, are so far not tested at all by experiment.
\item A comparison of the form factor discrepancy with the Bernauer phenomenological extraction allows for the possibility that the discrepancy might not stem from TPE alone.
\end{itemize}

We refer to \cite{Afanasev:2017gsk} for an in-depth review. 

The uncertainty in the resolution of the ratio puzzle jeopardizes the extraction of reliable form factor information, especially at the high $Q^2$ covered by the Jefferson Lab 12 GeV program. New data are needed to resolve the discrepancy. 

Both theory and phenomenological extractions predominantly predict a roughly proportional relationship of the TPE effect with $1-\varepsilon$ and a sub-linear increase with $Q^2$. $\varepsilon$ is the polarization of the virtual photon in the one-photon exchange approximation. Constraints on the non-linearities in the TPE effect are given in \cite{tvaskis06}. However, interaction rates drop sharply with smaller $\varepsilon$ and higher $Q^2$, corresponding to higher beam energies and larger electron scattering angles. The kinematic region of interest is easily accessible using the spectrometers in Hall A. As seen below, with the proposed positron source at Jefferson Lab, a clear measurement of the TPE could be made in two weeks.

\section*{Proposed Measurement}

In this proposal, we advance a new definitive measurement of the TPE effect that would be possible with a positron source at CEBAF. By alternately scattering positron and electron beams from a liquid hydrogen target and detecting the scattered lepton in the spectrometers available in Hall A, the magnitude of the TPE contribution between $Q^2$ values of 2 and 6~GeV$^2$, and at low $\varepsilon$, could be significantly constrained. With such a measurement, the question of whether or not TPE is at the heart of the ``proton form factor puzzle'' could be answered.

Hall A would provide a quick (two weeks) measurement of the TPE effect. By using the new Super BigBite Spectrometer (SBS) along with the upgraded BigBite Spectrometer (BB), and the existing High Resolution Spectrometer (HRS), we would be able to extend the measurement to a previously inaccessible kinematic region, $\varepsilon < 0.2$. The speed and precision of these measurements would be instrumental to addressing the ``form factor puzzle".

In addition to an unpolarized cross-section measurement, we would make use of a proton detection measurement. This approach minimizes false detector asymmetries, as detected trajectories are the same for both beam polarities. Further, the value of $Q^2$ is determined by the proton momentum alone, independent of beam energy, lepton kinematics and proton angle, which reduces the experimental systematic uncertainties. The $Q^2$ range is comparable to that of the lepton detection measurements proposed here, from 3 GeV$^2$ to 6 GeV$^2$.
 
 The measurement kinematics are chosen to determine the $\varepsilon$ dependence of the TPE contribution for fixed beam energy rather than fixed $Q^2$, comparable to the VEPP-3 and OLYMPUS experiments.

The initial state radiative corrections for either a traditional electron detection measurement, or a proton detection measurement are identical, as these corrections depend on the beam, not the detected particle. However, for a proton detection measurement there are a few advantages.
Primarily, the final state external bremsstrahlung for electrons is a significant correction, and is responsible for in- and out-scattering of electrons in the detector acceptance for a given $Q^2$. For the protons these corrections are significantly smaller. 

A similar proton-only measurement technique was used by Jefferson Lab experiments E01-001 and E05-017 to provide a more precise Rosenbluth extraction of the ratio $G_E/G_M$ for comparison to precise polarization measurements~\cite{qattan}. 

\begin{figure}[t!]
\centering
\includegraphics[width=0.86\linewidth]{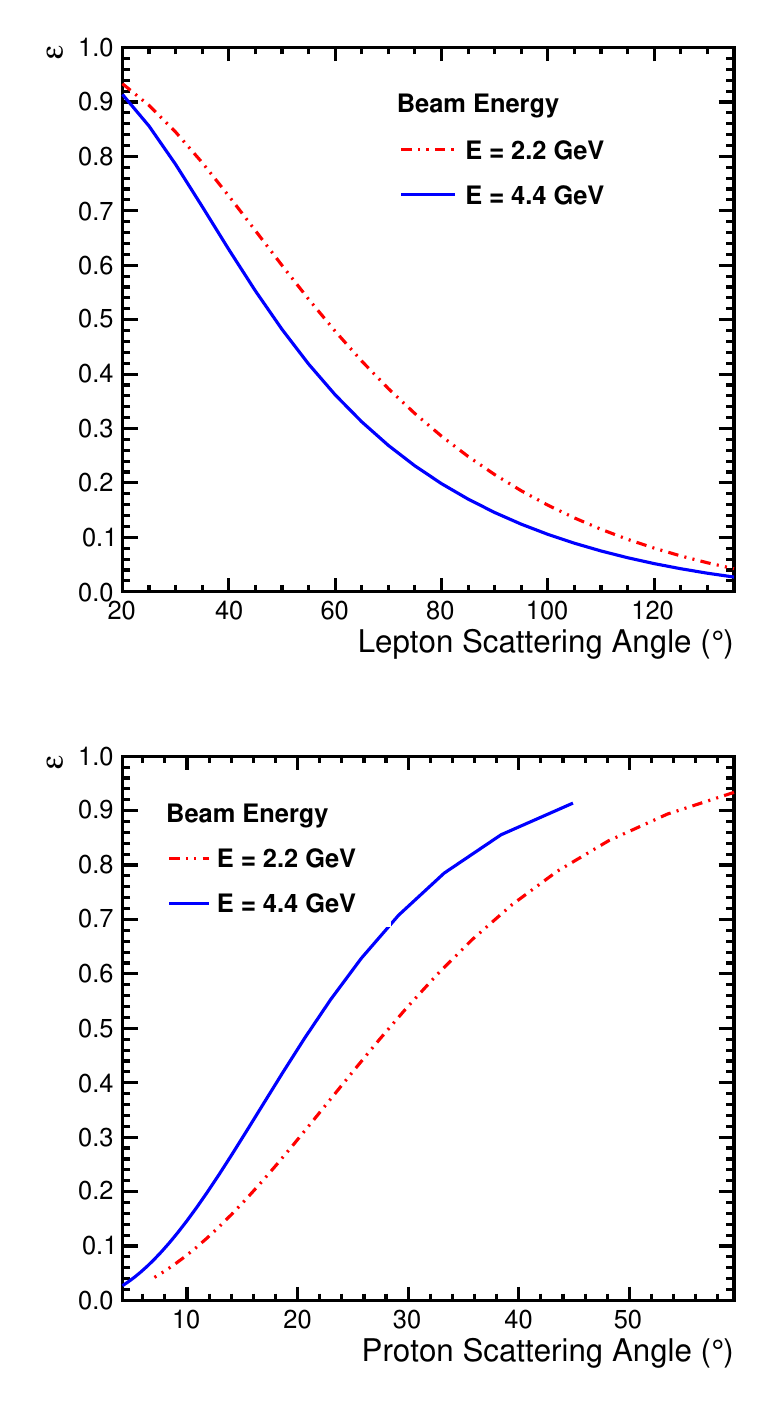}
\caption{$\varepsilon$ vs polar angle coverage for electron detection (top) and for proton detection (bottom) for the two proposed beam energies.}
\label{angle_reach}
\end{figure}

Figure~\ref{angle_reach} shows the angle coverage for both the electron (top) and for the
proton (bottom). There is a one-to-one correlation between the electron scattering angle and the proton recoil angle. For the kinematics of interest, $\varepsilon < 0.6$ and $Q^2 > 2$~GeV$^2$ for the chosen beam energies from 2.2 \& 4.4 GeV, nearly all of the electron scattering angles fall into a polar angle range from $40^\circ$  to $125^\circ$, and corresponding to the proton polar angle range from $8^\circ$ to $35^\circ$. These kinematics are most suitable for accessing the TPE contributions.

\begin{figure}[t!]
    \centering
    \includegraphics[width=\linewidth]{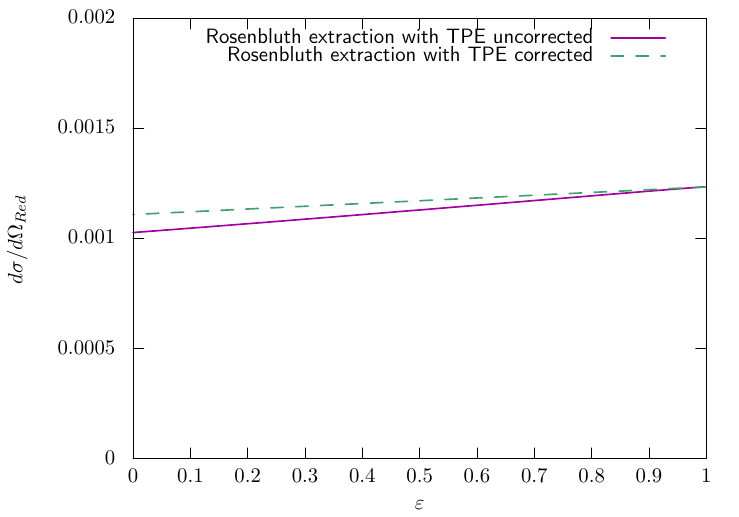}
    \caption{Example size of the TPE effect for $Q^2 = 6$ (GeV/c)$^2$. Here we plot the reduced cross section, $d\sigma/d\Omega_{Red} = \varepsilon G_E^2(Q^2) + \tau G_M^2(Q^2)$ against $\varepsilon$. Both curves appear linear in the scale of the figure, and are nearly identical at large $\varepsilon$. The TPE has a larger effect on the extraction of $G_E^2$, which is the slope of the reduced cross section. }
    \label{fig:tpe}
\end{figure}

It has been shown~\cite{arrington07} that the extraction of the high-$Q^2$ form factors is not limited by our understanding of the TPE contributions, as long as we assume that the Rosenbluth-Polarization discrepancy is explained entirely by TPE contributions. The proposed measurement would test this assumption, and also provide improved sensitivity to the overall size of the linear part of the TPE contribution that appears as a false contribution to $G_E$ when TPE contributions are neglected. An example of the size of the TPE effect is shown in Fig. \ref{fig:tpe}, where we plot the reduced cross section $d\sigma/d\Omega_{Red} = \varepsilon G_E^2(Q^2) + \tau G_M^2(Q^2)$. The measurement is also sensitive to non-linear contributions~\cite{tvaskis06} coming from TPE, and would provide improved sensitivity compared to existing electron measurements. More details are provided in Ref.~\cite{yurov17}.

\section*{Experimental Set-Up }
Here we discuss the setup for our measurement to be performed in Hall A. With the traditional Hall A HRS spectrometers the main kinematic consideration is the limited rate in each spectrometer at the chosen angles and beam energies. However, the large acceptance of BB and SBS allows measurements at very low values of $\varepsilon$ with excellent precision. 

\begin{figure}
    \centering
    \includegraphics[width=0.45\textwidth]{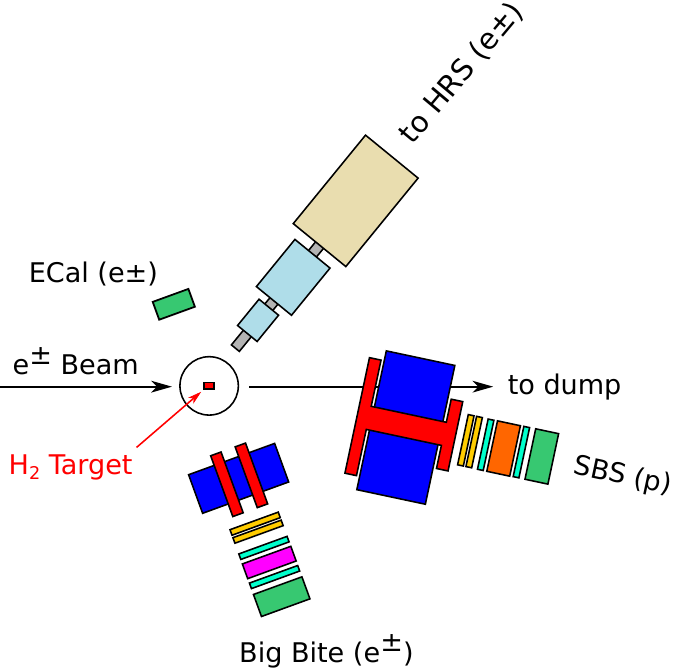}
    \caption{A not-to-scale schematic of the detector configuration for the proposed experiment. The particle type in parentheses indicate what will be detected by that spectrometer.}
    \label{fig:2gamma-exp}
\end{figure}

For the rate estimates and the kinematic coverage we have made a number of assumptions that are not overly stringent:
\begin{itemize}
\item Positron beam currents (unpolarized): $I_{e^+} \approx 1$~$\mu$A.  
\item Beam profile: $\sigma_x,~\sigma_y < 0.4$~mm.  
\item Polarization: not required, so phase space at the source maybe chosen for optimized yield and beam parameters.
\item Operate with a $\SI{10}{cm}$ liquid H$_2$ target and luminosity of $\SI{2.6}{\per\pico\barn\per\second}$ 
\item Use the HRS and BigBite for lepton ($e^+/e^-$) detection at $\theta_l = 40  - 120^\circ$.
\item Use the SBS for proton detection at $\theta_p = 6-15^\circ$ .
\item The SBS and BB will be placed on the same side of the beam line and will have a minimum opening angle of $58^\circ$. An ECal and the HRS will be placed on the opposite side with a minimum opening angle of $30^\circ$.
 \end{itemize}   
The Hall A configuration suitable for this experiment is shown in Fig.~\ref{fig:2gamma-exp}.

The HRS spectrometers are a pair of focusing magnetic spectrometers that is part of the standard spectrometer package in Hall A. One spectrometer is in the process of being decommissioned, but the second remains available for use in future physics measurements. The HRS has a small acceptance around 6 msr, and an excellent momentum resolution of 1$\times$10$^{-4}$. The angular resolution is 0.5 mrad in the horizontal direction, and 1 mrad in the vertical direction. Tracking and momentum measurements are performed by two vertical drift chambers, each with two planes of wires. Triggering and particle identification are performed by Cherenkov and time-of-flight scintillators \cite{Alcorn:2004}.

BigBite is a non-focusing magnetic spectrometer and is part of the standard spectrometer package in Hall A. Several existing proposals would plan to use BigBite as an electron arm \cite{Woj16,Woj19,Woj18,Ann17}. Though the specifics vary slightly, typically BigBite is envisioned employing GEM planes for tracking, a gas Cherenkov detector for electron ID, a lead glass calorimeter, and a scintillator hodoscope positioned between the pre-shower and shower layers of the calorimeter. For this proposal we will use a similar configuration. BigBite has an angular resolution of 4 mrad and a momentum resolution of 5 $\times$ 10$^{-3}$. The acceptance of BigBite is nearly 100 msr \cite{deLange:1998}.

SBS is a new spectrometer that is currently undergoing commissioning in Hall A at Jefferson Lab. SBS is designed to operate with a solid angle of up to 70 msr and at the highest luminosity possible available from CEBAF. Rather than traditional drift chambers, SBS utilizes GEM chambers capable of handling luminosity on the order of 10$^{39}$ cm$^{-2}$/s.
The several approved SBS experiments (e.g., \cite{Ann17,Cis08,Woj16,Woj18,Woj19}) use a variety of detector configurations. For the measurement proposed here, the detector stack will consist of a front tracker, followed by time-of-flight scintillators, and a Cherenkov detector capable of separating protons from other positively charged hadrons. The upcoming ``Transversity Experiment'' \cite{Woj18} will employ the ring-imaging Cherenkov (RICH) detector from the HERMES experiment to perform $\pi^+/K^+/p$ separation. This would be more than sufficient for the measurement proposed here. Alternatively, a new aerogel threshold Cherenkov counter would be a simpler alternative.

SBS is capable of utilizing larger targets than are typical at Jefferson Lab, up to 40 cm. For the kinematics in this proposal, SBS has an angular resolution of $\approx$0.68 mrad, and a momentum resolution of $\approx$4$\times$10$^{-3}$. SBS is capable of reaching extremely small scattering angles, down to 3.5$^{\circ}$, although the acceptance decreases in proximity to the beam pipe \cite{deJager:2010}. For this proposal SBS will have an acceptance between 12 and 40 msr.

SBS does not have the momentum resolution to fully eliminate inelastic background on its own. For that reason, an ECal is necessary to tag leptons coincident with the proton detected in SBS. The combined coincidence measurement will reduce the inelastic background to negligible levels. Such an approach has been used previously at Jefferson Lab, e.g., the BigCal detector in the GEp-III experiment in Hall C~\cite{puckett10,puckett17} and the ECal planned for the upcoming GEp-V experiment using SBS~\cite{Cis08}. The calorimeter will need to be large enough to match the acceptance of SBS. As the experimental floor will be crowded with other spectrometers, detailed simulations would need to be performed to precisely determine the location where the calorimeter would be placed.

The proposed measurement program for Hall A is listed in Tab.\ \ref{halla2}. While these measurements could provide precise measurements over a range of $\varepsilon$ values in a short run period, they cover a limited range of beam energies. Because the measurements suffer from the same beam-related systematics, they would benefit from rapid change-over between positrons and electrons. The luminosity would best be measured by beam current monitors. Since existing monitors are not optimized for a 1 $\mu$A current, there would necessarily be an upgrade to these detectors as a part of the positron source installation.

Proton detection is also beneficial in making precise comparisons of electron and positron scattering. As the SBS will be used to only detect protons throughout the measurement, there is no need to change the polarity of the magnetic field.

The general measurements would be similar to the E05-017 experiment \cite{Arr06}, with the exception of using a low intensity positron beam and alternating with a similar electron beam. Assuming a 1~$\mu$A positron beam and the 10~cm LH2 target used in E05-017, a 12 day run could provide measurements with sub-percent statistical uncertainties from 2.2 \& 4.4~GeV beam energies, yielding total uncertainties comparable to the electron beam measurements. An additional 2 days would be requested for empty target measurements.

\begin{table*}[h]
\newcommand{\PERcen}[1]{\multicolumn{1}{|c|}{#1}}
  \begin{center}
    \caption{\label{halla2}The proposed measurement plan in Hall A. The total proposed measurement time is 12 days plus 2 additional days for empty target measurements. The angles listed are the central angles of the HRS, BigBite, and SBS, respectively. For the SBS angles, the number in parentheses is the corresponding lepton quantity measured by the ECal.}
    \begin{tabular}{|l|ccc|ccc|ccc|}
    \hline
    $E_{beam}$          & \multicolumn{3}{c|}{2.2 GeV}& \multicolumn{3}{c|}{2.2 GeV} &  \multicolumn{3}{c|}{4.4 GeV} \\ \cline{3-4}
    \hline
    Spectrometer & HRS & BigBite & SBS(ECal)  & HRS & BigBite & SBS(ECal)  & HRS & BigBite & SBS(ECal) \\
    Spec. Angles ($^\circ$) & 50 & 70 & 12 (110) & 80 & 120 & 6.2 (140) & 40 & 80 & 14 (70)\\
    $Q^2$ [(GeV/$c$)$^2$] & 1.9 & 2.5 & 3.1 & 2.7 & 3.2 & 3.3 & 4.3 & 6.6 & 6.2 \\ 
    $\epsilon$ & 0.59 & 0.37 & 0.11 & 0.28 & 0.08 & 0.03 & 0.62 & 0.19 & 0.26 \\
    Momentum (GeV/$c$) & 1.197 & 0.865 & 2.4 (0.530)  & 0.75 & 0.49 & 2.54 (0.43)  & 2.10 & 0.90 & 4.16 (1.08) \\
    Solid Angle (msr) & 6 & 100 & 30 & 6 & 100 & 12 & 6 & 100 & 40 \\
    Time [day/species] & \multicolumn{3}{c|}{1} & \multicolumn{3}{c|}{2} & \multicolumn{3}{c|}{3}\\
    \hline
    \end{tabular}
  \end{center}
\end{table*}

\begin{figure}[t!]
  \centerline{\includegraphics[width=\linewidth]{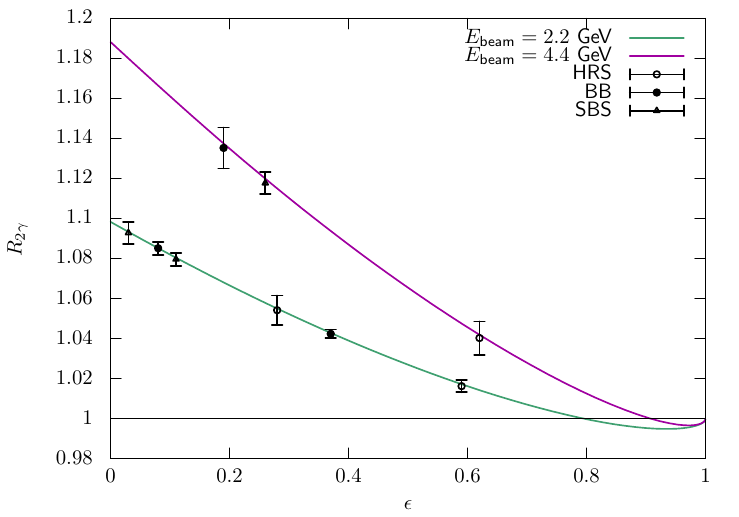}  }
  \centerline{\includegraphics[width=\linewidth]{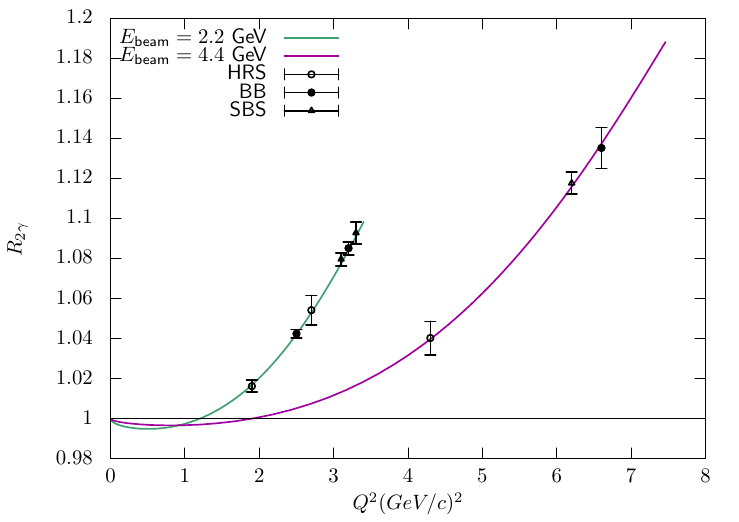}  }
  \caption{\label{figerrjlab}Predicted effect size from the Bernauer phenomenological TPE parameterization and estimated statistical errors in the region of interest in Hall A.}
\end{figure}

Figure \ref{figerrjlab} shows the estimated errors and predicted effect size as a function of $\varepsilon$ and $Q^2$. The statistical uncertainty was obtained by calculating the cross section at the central spectrometer angle, multiplied by the spectrometer acceptance, and calculating the number of events measured in 1, 2, or 3 days of running. 

The proposed experiment will yield a high-impact measurement with two weeks of allocated beam time. Even in the case the final positron beam current is lower than assumed here, the experiment remains feasible.

\section*{Systematic Uncertainties}

At the most forward SBS angles the difference in momentum between the elastic lepton-proton peak and the inelastic threshold for single pion production are separated by $<$ .5\%. This difference is not cleanly resolvable in the SBS. In order to suppress this background, we will measure the coincident scattered lepton in the ECal. A coincidence measurement of this nature will allow for the selection of true elastic lepton-proton scattering events.

The main benefit to measure both lepton species in the same setup closely together in time is the cancellation of many systematics which would affect the result if data of a new positron scattering measurement is compared to existing electron scattering data. For example, one can put tighter limits on the change of detector efficiency and acceptance between the two measurements if they are close together in time, or optimally, interleaved. 

To make use of these cancellations, it is paramount that the species switch-over can happen in a reasonably short time frame ($<1$~day) to keep the accelerator and detector setup stable. For the higher beam energies, where the measurement time is longer, it would be ideal if the species could be switched several times during the data taking period.

For the ratio, only relative effects between the species types are relevant; the absolute luminosity, detector efficiency, etc.\ cancel. Of special concern here is the luminosity. While an absolute luminosity is not needed, a precise determination of the species-relative luminosity is crucial. Fortunately, the luminosity can easily be monitored to sub-percent accuracy based on beam current measurements and monitoring the target density. The standard Hall A cryotarget is designed to withstand a 100~$\mu$A beam with no more than 1\% reduction in density, vastly more strenuous conditions than in this proposal. The beam current monitors in Hall A are conservatively estimated to have 1\% accuracy. This system would likely need to be upgraded to cope with beam currents below 1~$\mu$A.

To keep the beam properties as similar as possible, the electron beam should not be generated by the usual high quality source, but employ the same process as the positrons. This will help minimize any differences in effects such as beam power on the target, beam dispersion, etc.

An additional source of systematic uncertainty is positron annihilation in flight, however estimates put this at a level well below relevance. Further, there will be some positron background in the SBS, but none from electrons. This will be a small difference in background between the two beam polarities. SBS should have sufficient discriminating power in the hadron calorimeter to differentiate these particles.

Finally, it will be necessary to perform a target background subtraction via an empty target measurement. It is difficult to estimate the statistics that will need to be collected without knowing the detailed properties of the electron/positron beam that will be built at Jefferson Lab, but we conservatively estimate 2 days of data taking total. This time would be divided between both beam energies and all detector configurations as necessary.

\section*{Conclusion}

Despite recent measurements of the $e^+p/e^-p$ cross section ratio, the proton's form factor discrepancy has not been conclusively resolved, and new measurements at higher momentum transfer are needed. With a positron source at CEBAF, the enormous capabilities of the Hall A spectrometers can be brought to bear on this problem and provide a wealth of new data over a widely important kinematic range. 

Using the existing and in-development spectrometers in Hall A, our proposed measurement could be completed with a typical spectrometer configuration following the construction of a positron source. The measurement using unpolarized lepton-proton scattering allows for a comparison with existing electron scattering data, while extending the search for TPE contributions to the proton form factors.

The additional measured points utilizing proton-only detection in SBS allow us to  extend the data to higher $Q^2$ with small statistical uncertainties and reduced systematics, as beam-related effects are suppressed.

The data that the proposed experiment could provide covers the region of interest for determining the impact of TPE on the form factor puzzle and can be performed quickly. This experiment will be able to map out the transition between the regions of validity for hadronic and partonic models of hard TPE, and make definitive statements about the nature of the proton form factor discrepancy.

\section*{Acknowledgements}
This work was supported in part by the National Science Foundation grant number 2012114.




\end{document}